\documentclass[pra,twocolumn,showkeys,preprintnumbers,amsmath,amssymb]{revtex4}
\makeatletter
\usepackage[dvips]{graphicx}	

\usepackage{amsmath}
\usepackage{amssymb}
\usepackage{amsbsy}
\usepackage{color}
\setcitestyle{square}
\usepackage{epstopdf}
\DeclareGraphicsExtensions{.pdf,.eps,.png,.jpg,.mps}

\begin{document}

\title{
Controlling spin polarization of gapless states  in defected trilayer graphene \\ with a gate voltage }
\author{ W. Jask\'olski} 
\email{wj@fizyka.umk.pl}
\affiliation{Institute of Physics, Faculty of Physics, Astronomy and Informatics, Nicolaus Copernicus University in Torun, Grudziadzka 5, 87-100 Toru\'n, Poland}

\begin{abstract}
Trilayer graphene exhibits valley-protected gapless states when the stacking order changes from ABC to CBA and a gate voltage is applied to outer layers. Some of these states survive strong distortions of the trilayer. For example, they persist when the outer layers are partially devoid yielding a system of two trilayers of different stacking order connected by a strip of a single graphene layer. Here we investigate how these states respond to another perturbation, i.e., the presence of magnetic defects, which we model as $\pi$-vacancies. We show that the gap states hybridize with the defect states and strongly spin-split. More importantly, it is demonstrated that by changing the gate voltage value one can change the spin density of the gap states and the corresponding currents at the Fermi level.  
\end{abstract}

\keywords{trilayer graphene; topological states; defects in graphene} 

%\date{\today}

\maketitle

\section{Introduction}

Multilayer graphene is attracting still attention due to strongly correlated states and superconductivity reported both, in the systems with twisted layers \cite{Jarillo2018, Chen_Nature_2019, Chittari_PRL_2019, Yin_PRB_2020, Liu_Nature_2020,Shen_Nat_Phys_2020} and more recently in non-twisted Bernal stacked bilayer and rhombohedral trilayer graphene under special conditions \cite{Zhou_Nature_2021,Zhou_Science_2022,Pantaleon_NRP_2023}.  Multilayers are attracting also interest in electronic applications due to the opening of the tunable energy gap when the systems are gated \cite{Ohta_2006, Castro_2007,Oostinga_2008,
Zhang_Nature_2009,Peeters1,Peeters2,Szafranek_2010,Padilha_2011,Schwierz_2010,Lin_2008,Choi_2010,Santos_2012,Zhang_transistor_2018}.

Another interesting property of gated Bernal stacked bilayer or rhombohedral trilayer is the appearance of valley-protected gap states of topological character when the stacking order changes from AB to BA in bilayer or from ABC to CBA in trilayer \cite{Yin_NC_2016,
Vaezi_2013,Alden_2013,San_Jose_2014,Jaskolski_2020}. The stacking order change occurs usually when one of the layers is stretched, corrugated, or delaminated \cite{Ju_Nature_2015,Pelc_2015,Lane_2018,Anderson_PRB_2022}.
The gapless states are important since they provide one-dimensional conducting channels at the Fermi level ($E_F$) along the stacking domain walls. An important feature of these states is their robustness against structural deformations of multilayers. They largely survive in the presence of atomic-scale defects \cite{Jaskolski_2016,Jaskolski_RSC_2019} which introduce defect states into the energy gap, and thus may disrupt topological states. 
Some of them persist even when the multilayer is partially stripped of one or two layers \cite{Jaskolski_2019,Jaskolski_2021,Jaskolski_2023}. 

In this work, we consider strongly defected trilayer graphene, i.e., devoid of outer layers in the region of the stacking domain wall.
This system was recently studied in Ref. \cite{Jaskolski_2023}, 
but here we add another perturbation, i.e., $\pi$-vacancy defects.  
Since vacancies in graphene lead to the appearance of localized states and magnetic moments, we use them here as simple models of magnetic defects \cite{Pereira_2006,Palacios_2008,Lopez_2009,
Kusakabe_book,Jaskolski_RSC_2019}. Our aim is to investigate how such defects influence gapless states, in particular how they remove spin degeneracy of these states, what may be important for applications in spintronic devices. We find that the spin polarization and spin density of the gap states and the corresponding one-dimensional currents at the Fermi level depend strongly on the value of gate voltage applied to outer layers of the trilayer.

\section{System description and method of calculation}

\begin{figure}[ht]
\centering
\includegraphics[width=\columnwidth]{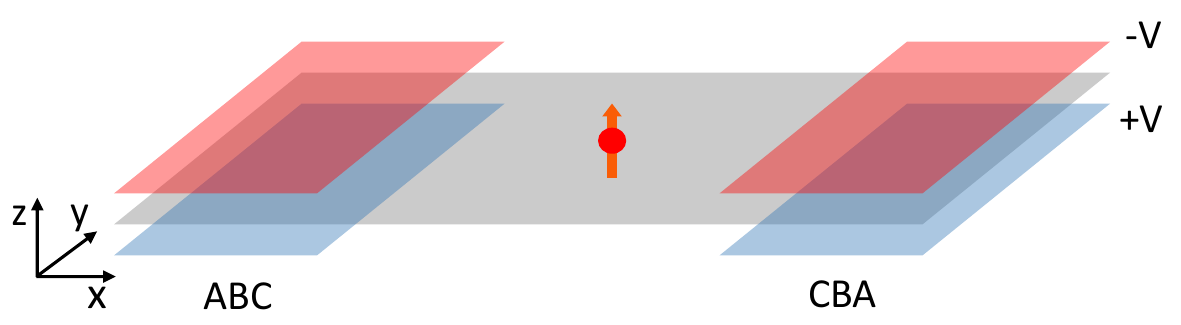}
\caption{\label{fig:first}
Schematic representation of the investigated system.
The left and right trilayers have ABC and CBA arrangement of layers, respectively. They are connected by a strip of single graphene layer, i.e., the middle layer of trilayers. The system extends to infinity in the $x$ (armchair) and $y$ (zigzag) directions but is fully periodic only in the $y$ direction.
A single vacancy representing magnetic impurity, marked as a red dot and arrow, is located periodically along the $y$ direction in the region of the single graphene layer. 
}
 \end{figure}

The system under investigation is schematically shown in Fig. \ref{fig:first}. It consists of two graphene trilayers connected by a single layer strip. The stacking order of the trilayers on the left and right sides is ABC and CBA, respectively. Therefore, the system can be also seen as a trilayer graphene with ABC/CBA stacking domain wall and the outer layers devoid in the region of the domain wall. It is worth noticing that because both outer layers are torn and pulled apart, the stacking domain wall area extends into the central region, i.e., into the single-layer strip.

%% FIGURE %%%

The system is infinite in both, the $x$ (armchair) and the $y$ (zigzag) directions, but is fully periodic in the zigzag ($y$) direction. The width of the system unit cell in the periodic ($y$) direction is $W_y=4$, measured as the number of graphene unit cells along this direction.
The width of the central region along the $x$ (armchair) direction, i.e., the width of the single graphene strip connecting two trilayers, is taken as $W_C=4$, measured in the same units. Each unit cell of the system contains a single vacancy (as shown in Fig. \ref{fig:first}), that in bipartite systems introduces magnetic moment and thus can model magnetic defect \cite{Kusakabe_book}.

It is important to note that although we study a model of uniform distortion of the trilayer graphene (i.e., the single layer strip has a  constant width and vacancies are periodically distributed)
the robustness of gapless states to different perturbations allows us to assume that the obtained results and conclusions could be applied also to systems not so uniformly perturbed.

We use in the calculations a one-orbital $\pi$-electron tight-binding approximation (TB). This approach has proved to properly model the electronic properties of graphene systems around the  Fermi energy. The electron-electron interaction is taken into account by including a Hubbard term, which is adequate for the description of spin and magnetic effects in graphene within the TB model \cite{Kusakabe_book}. The Hubbard Hamiltonian in a mean-field approximation is 
$$ H=t_{i/e}\sum_{{\langle}i,j{\rangle},\sigma} c_{i\sigma}^{\dagger}c_{j\sigma} + H.c. + U\sum_{i}(n_{i\uparrow}{\langle}
n_{i\downarrow}{\rangle}+{\langle} 
n_{i\uparrow}{\rangle}n_{i\downarrow}),$$
where $c_{i\sigma}^{\dagger}$ ($c_{i\sigma}$) are the creation and (annihilation) operators for electrons with spin $\sigma$ at site $i$; the index $i$ goes over all the nodes in the unit cell; the summation
${\langle}i,j{\rangle}$ is restricted to nearest neighbors; the arrows indicate spin-up and spin-down $\sigma$ states; and 
${\langle}n_{i\sigma}{\rangle}={\langle}c_{i\sigma}^{\dagger}c_{i\sigma} {\rangle}$ is spin-resolved density at site $i$. The first term in $H$ is the TB Hamiltonian, while the last one represents the on-site Coulomb repulsion. Intra-layer and inter-layer hopping parameters $t_i=2.7$ eV and $t_e = 0.27$ eV are used, respectively \cite{Castro_2007,Ohta_2006}, the on-site Coulomb repulsion parameter $U$ is set equal to 2.8 eV \cite{ Gunlycke_2007,Jaskolski_divac_2015,Jaskolski_RSC_2019}. 

To calculate the local density of states (LDOS) we use the Green function matching technique \cite{Nardelli_1999}. The Hamiltonians $H_C$, $H_L$ and $H_R$ of the central region (i.e., single layer square $[W_C \times W_y ]$ shown in Fig. \ref{fig:first}) and of the left and right trilayers are calculated self-consistently since the densities ${\langle}n_{i\sigma}{\rangle}$ depend on the eigenvalues of the Hamiltonians. Knowing the $H_{L/R/C}$ Hamiltonians, the transfer matrix technique \cite{FGM} is employed to find the Green function $G_C$ of the central region, and the corresponding LDOS is calculated as 
LDOS=$-(\frac{1}{\pi}){\rm{Tr}}G_C$ \cite{Datta}. Since the system is periodic in the $y$ (zigzag) direction, the LDOS is $k$-dependent, where $k$ is the wave vector corresponding to this periodicity. Therefore, the entire procedure for finding $H_{L/R/C}$, $G_C$ and LDOS has to be performed for each $k$ value in the Brillouin Zone, i.e., from $k=0$ to $k=\pi /a$, where $a=W_y$.

%% FIGURE %%%
\begin{figure}[ht]
\centering
\includegraphics[width=\columnwidth]{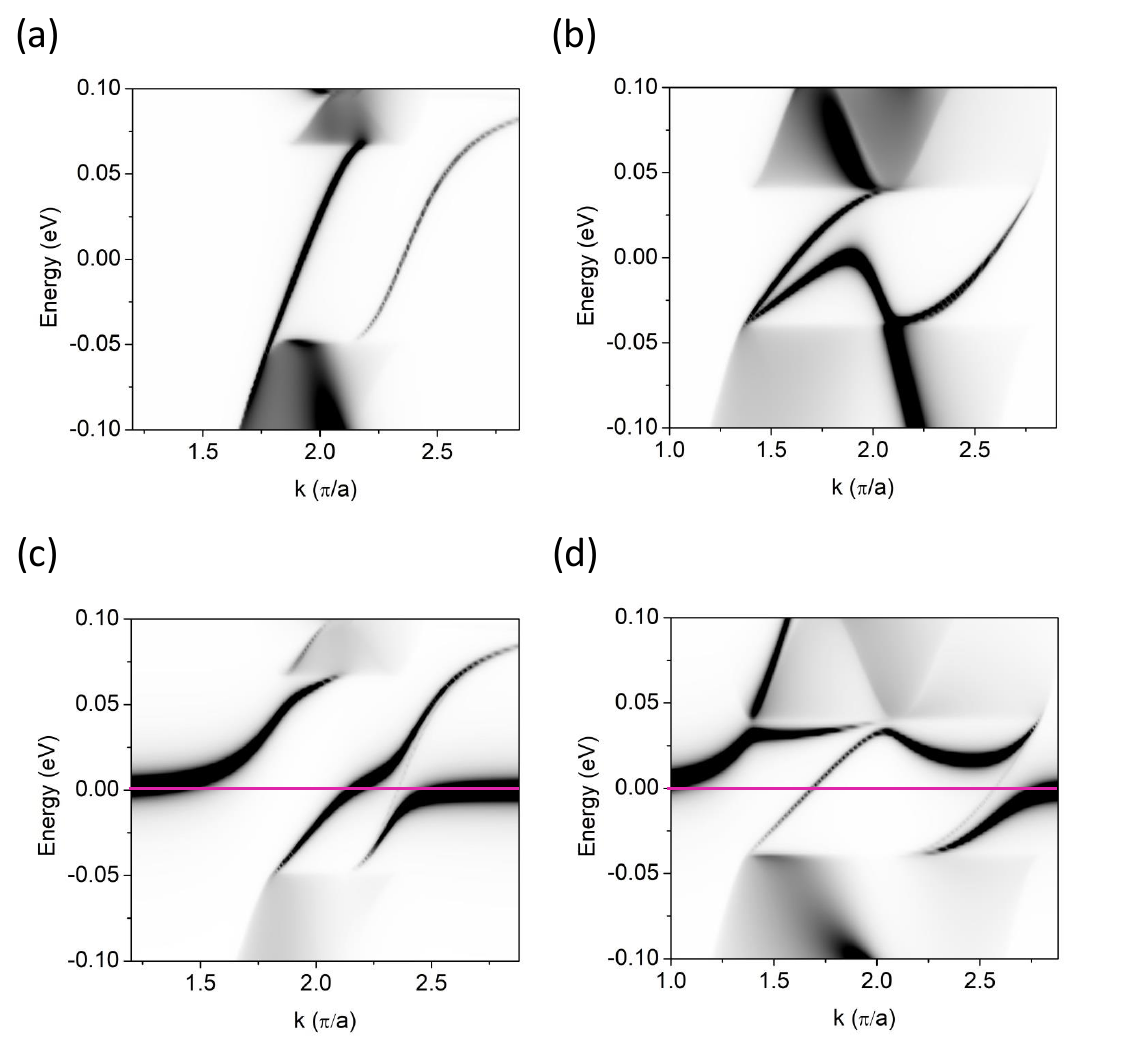}
\caption{\label{fig:second}
LDOS visualized close to the energy cone, i.e., near the Fermi level ($E=0$)  and for $k$ around $\frac{2}{3} \pi $. (a) and (c) $V=0.1$ eV, (b) and (d) $V=0.4$ eV. 
Upper panels: LDOS calculated for the case without vacancies. 
Lower panels: LDOS calculated for system with vacancies, but without Coulomb repulsion, i.e., setting $U=0$. Pink solid line marks the position of the defect states for the case of gated trilayer without stacking order change.
%
%Although the LDOS is calculated in the central single-layer 
%strip of graphene it contains also some traces of the electronic %structure (the energy band continua and the gap) 
%of the neighboring gated trilayers. 
}
 \end{figure}

%% FIGURE %%%
\begin{figure}[ht]
\centering
\includegraphics[width=\columnwidth]{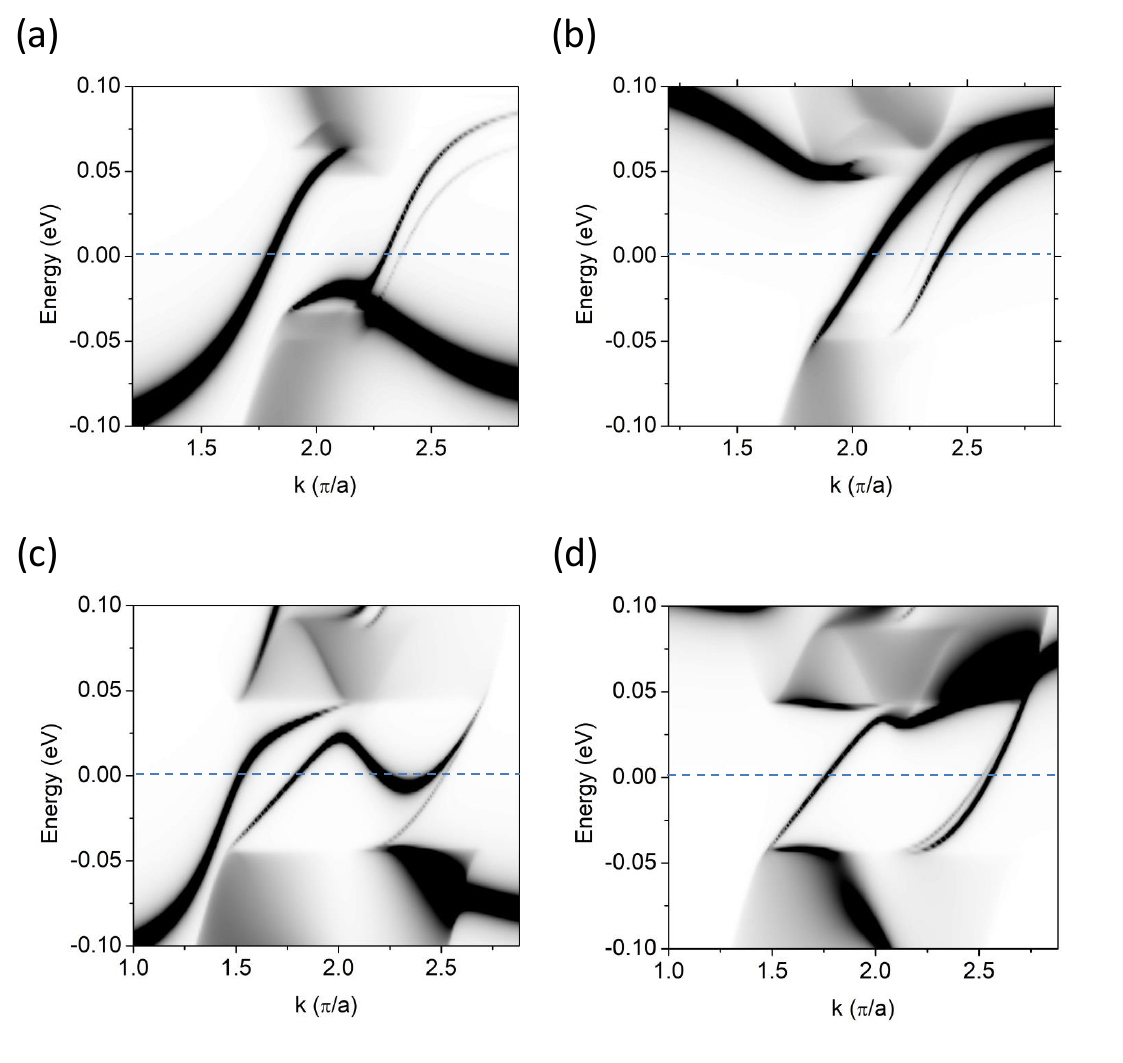}
\caption{\label{fig:third}
Spin-resolved LDOS calculated for the case with vacancies present in the central region of the system.  (a) and (b) $V=0.1$ eV, (c) and (d) $V=0.4$ eV. (a) and (c) spin down, (b) and (d) spin up. The Fermi level is marked by a dashed line.
}
 \end{figure}

\section{Results and discussion}

We consider two values of the gate voltage $\pm V$  applied to the outer layers, namely $V=0.1$ eV and $V=0.4$ eV. As shown in Ref. \cite{Jaskolski_2023}, different values of $V$, larger or smaller than $t_e$, lead to different number and behavior of the gap states in trilayer graphene partially devoid of the outer layers. This is visualized in Fig. \ref{fig:second} (a) and (b), where the results for the case with no vacancies are presented. The LDOS is calculated in the central part of the system and only LDOS close to the energy cone, i.e., close to $k= \frac{2}{3}\pi$ and the Fermi energy ($E=0$), is visualized. 
Although the LDOS is calculated in the region of the single graphene layer, one can clearly identify gap states characteristic for multilayer graphene with stacking order change. The LDOS shows also some traces of the electronic structure of the neighboring gated trilayers, i.e., the band continua and the energy gap. 

For $V=0.1$ eV, two states of similar and monotonic behavior of $E(k)$ are present in the energy gap. As shown in Ref. \cite{Jaskolski_2023} there are in fact three gap states, since the right one is doubly degenerate in energy.  This right and degenerate pair of the gap states couples to a pair of degenerate zigzag edge states localized in the lower half-layers (blue in Fig. \ref{fig:first}) of the left and right trilayers \cite{nota_zigzag}. 
For $V=0.4$ eV, one of the gap states changes twice the slope of $E(k)$, but as it was explained in Ref. \cite{Jaskolski_2023} the rightmost part of this state overlaps with the third gap state. 

We now analyze the influence of the vacancy defects. 
When the Coulomb interaction is not allowed, i.e., when we set $U=0$ in the Hubbard Hamiltonian, the vacancies introduce defect state at $E=0$ (no gate is applied to the middle layer), which strongly interacts and hybridizes with the gap states. 
This is visualized in Fig. \ref{fig:second} (c) and (d) for $V=0.1$ eV and $V=0.4$ eV, respectively.

%% FIGURE %%%
\begin{figure}[ht]
\centering
\includegraphics[width=\columnwidth]{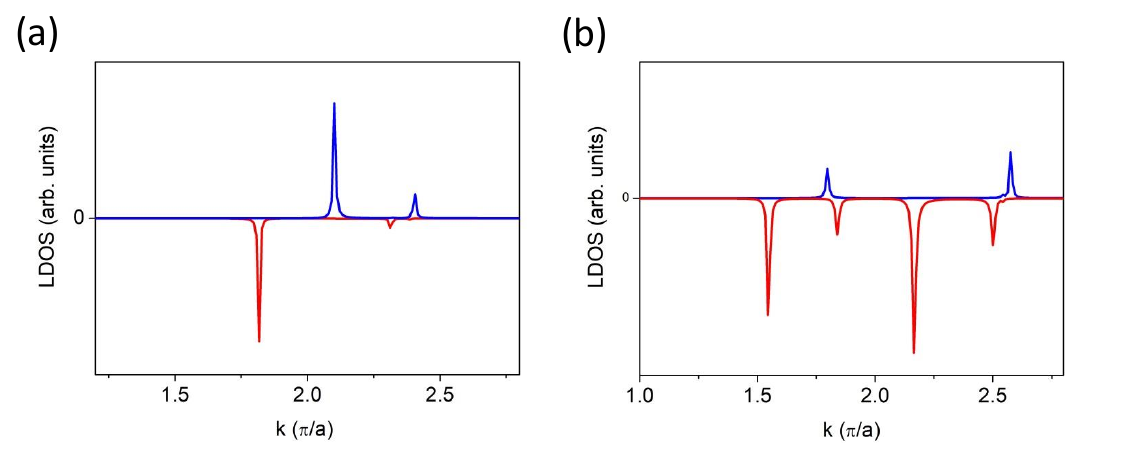}
\caption{\label{fig:fourth}
Spin-resolved LDOS at the Fermi level. (a) $V=0.1$ eV, (b) $V=0.4$ eV. Spin-down and spin-up LDOS are marked in red and blue, respectively.
}
 \end{figure}

All these states are spin-degenerate so when the Coulomb interaction is switched on they strongly spin-split. 
Figs. \ref{fig:third} (a) and (b) show the spin-down and spin-up gap states, respectively, calculated for the case of $V=0.1$ eV. 
Two gap states connecting the valence and conduction band areas are clearly visible for both spin polarization. The spin-splitting of the left gap state is larger than the right one because, as demonstrated in Ref. \cite{Jaskolski_2023}, this state is localized mainly in the single layer region and therefore is more affected by  vacancies, which are also located in this layer. 

The spin-down and spin-up states for the case of $V=0.4$ eV are shown in panels (c) and (d) of Fig. \ref{fig:third}, respectively. 
Of two spin-down states,  the right one follows the behavior of the right state of the vacancy-free case, while both spin-up states show monotonic dependence of their energies vs. wave vector, $E(k)$, almost in the entire energy gap. The picture of spin-splitting of the gap states is more complex than in the $V=0.1$ eV case: the right spin-down state changes twice the slope of $E(k)$ and thus it crosses three times the Fermi level. It means that the density of the occupied spin-down gap states at $E_F$ is much higher than the density of the spin-up states. This is visualized in Fig. \ref{fig:fourth} (b), where the spin-down and spin-up LDOS at the Fermi level is presented.  For comparison, the LDOS at $E_F$  of the $V=0.1$ eV case is shown in panel (a) of this Figure. In this case, the spin-down and spin-up densities are almost the same.

The gap states at $E_F$ can curry one-dimensional and spin-polarized currents along the $y$ direction when the system is additionally biased in this direction. The presented results show that by changing the value of the gate voltage one can change the density of spin-polarized gap states and the corresponding currents at the Fermi level. This is the main message of this work: a slight change of the gate voltage from 0.1 eV to 0.4 eV can serve as a switch from spin-unpolarized current to a polarized one.  

The behavior of gap states away from the cone is governed by the defect state, which for that values of $k$ strongly splits into sin-down and spin-up states with energies below and above the cone, respectively. Since most of the vacancy-hybridized spin-up bands lies above the Fermi level and is unoccupied, the magnetic moment (estimated from Fig. \ref{fig:fourth}) of the central region is about 0.9 $\mu_B$ and 0.6 $\mu_B$ for $V=0.1$ eV and $V=0.4$ eV, respectively.

A comment is required about the barely visible gap state that appears at the right side of the energy cone in all panels of Fig. \ref{fig:third}. 
This is the above mentioned third gap state of the right degenerate pair of the vacancy-free case. This state is localized almost exclusively in the lower layers and on the sublattice defined by the zigzag edge nodes of the lower left half-layer. This sublattice does not couple to the vacancy-defined sublattice of the middle layer (see Ref. \cite{nota_zigzag}). For this reason its LDOS in the middle layer is very small, it does not hybridize with the vacancy state and almost does not spin-split.

\section{Conclusions}

We have studied the electronic structure of defected gated trilayer graphene with stacking order change of the layers from ABC to CBA. The defect comes down to the partial removal of the outer layers in the region of the stacking domain wall and the inclusion of vacancies, which mimic the presence of magnetic defects. We have investigated the role of vacancies in the spin-splitting of gapless states. In particular, we have checked how this splitting, and thus the spin-resolved density of gapless states at the Fermi level, depends on the value of voltage applied to the outer layers. 

The calculations have been performed within the tight binding approximation and the Hubbard model. The surface Green function matching technique has been used to calculate the local density of states in the defected region.

We have shown that gapless states present in the trilayer system due to the stacking order change are strongly affected by the vacancy defects.
The interaction of the vacancy state with gapless states and their spin-splitting depends strongly on the value of the gate voltage. When the applied voltage is lower than the interlayer hopping energy $t_e$, the pair of the resulting spin-down and spin-up gap states have similar and uniform slope of $E(k)$, yielding zero net spin density at the Fermi level. In contrast, when the gate voltage is higher than $t_e$, one of the spin-down states has a more complex curvature of $E(K)$ than its spin-up counterpart. As a result, one spin density of the gap states dominates at the Fermi level. Therefore, the one-dimensional currents corresponding to the gap states are also spin-polarized, the effect of potential application in spintronics based on multilayer graphene systems.


\begin{thebibliography}{99}
% superconductivity twisted
\bibitem{Jarillo2018} Y. Cao, V. Fatemi, S. Fang, K. Watanabe, T. Taniguchi, E. Kaxiras, and P. Jarillo-Herrero, Nature {\bf 556}, 43 (2018).
\bibitem{Chen_Nature_2019} E. Chen, A. L. Sharpe, P. Gallagher, I. T. Rosen, E. J. Fox, L. Jiang, B. Lyou, H. Li, K. Watanabe, T. Taniguchi, J. Jung, Z. Shi, D. Goldhaber-Gordon, Y. Zhang, and F. Weng, Nature {\bf 572}, 215 (2019).
\bibitem{Chittari_PRL_2019} B. L. Chittari, G. Chen, Y. Zhang, F. Wang, and J. Jung, Phys. Rev. Lett. {\bf 122}, 016401 (2019).
\bibitem{Yin_PRB_2020} L.-J. Yin, L.-Z. Yang, L. Zhang, Q. Wu, X. Fu, L.-H. Tong, G. Yang, Y. Tian, L. Zhang, and Z. Qin, Phys. Rev. B {\bf 102}, 241403(R) (2020).
\bibitem{Liu_Nature_2020} X. Liu, Z. Hao, E. Khalaf, J. Y. Lee, Y. Ronen, H. Yoo, D. H. Najafabadi, K. Watanabe, T. Taniguchi, A. Vishwanath, and P. Kim, Nature {\bf583}, 221 (2020).
\bibitem{Shen_Nat_Phys_2020} Ch. Shen, Y. Chu, Q. Wu, N. Li, S. Wang, Y. Zhao, J. Tang, J. Liu, J. Tian,K. Watanabe, T. Taniguchi, R. Yang, Z. Y. Meng, D. Shi, O. V. Yazyev, and G. Zhang, Nat. Phys. {\bf 16}, 520 (2020). 
% superconductivity non-twisted – nowe
\bibitem{Zhou_Nature_2021} H. Zhou, T. Xie, T. Taniguchi, K. Watanabe, and A. F. Young, Nature {\bf 598}, 434 (2021).
\bibitem{Zhou_Science_2022} H. Zhou, L. Holleis, Y. Saito, L. Cohen, W. Huynh, C. L. Patterson, F. Yang, T. Taniguchi, K. Watanabe, and A. F. Young, Science {\bf 375}, 774 (2022). 
\bibitem{Pantaleon_NRP_2023} P. A. Pantaleon, A. Jimeno-Pozo, H. Sainz-Cruz, V. T. Phong, T. Cea, F. Guinea, Nat. Rev. Phys. {\bf 5}, 304 (2023).
\bibitem{Ohta_2006} T. Ohta, A. Bostwick, T. Seyller, K. Horn, and E. Rotenberg, Science {\bf 313}, 951 (2006).
\bibitem{Castro_2007} E. V. Castro, K. S. Novoselov, S. V. Morozov, N. M. R. Peres, J. M. B. L. dos Santos, J. Nilsson, F. Guinea, A. K. Geim, and A. H. C. Neto, Phys. Rev. Lett. {\bf 99}, 216802 (2007).
\bibitem{Oostinga_2008} J. B. Oostinga, H. B. Heersche, X. Liu, A. F. Marpurgo, and L. M. K. Vandersypen, Nat. Mater. {\bf 7}, 151 (2008).
\bibitem{Zhang_Nature_2009} Y. Zhang, T.-T. Tang, C. Girit, Z. Hao, M. C. Martin, A. Zettl, M. F. Crommie, Y. R. Shen, and F. Wang, Nature (London) {\bf 459}, 820 (2009).
\bibitem{Peeters1} A. A Avetisyan, B. Partoens, and F. M. Peeters, Phys. Rev. B {\bf 80}, 195401 (2009).
\bibitem{Peeters2} A. A Avetisyan, B. Partoens, and F. M. Peeters, Phys. Rev. B {\bf 81}, 115432 (2010).
\bibitem{Szafranek_2010} B. N. Szafranek, D. Schall, M. Otto, D. Neumaier, and H. Kurz, App. Phys. Lett. {\bf 96}, 112103 (2010).
\bibitem{Padilha_2011} J. E. Padilha, M. P. Lima, A. J. R. da Silva, and A. Fazzio, Phys. Rev. B {\bf 84}, 113412 (2011).
\bibitem{Schwierz_2010} F. Schwierz, Nat. Nanotechnol. {\bf 5}, 487 (2010).
\bibitem{Lin_2008} Y.-M. Lin and P. Avouris, Nano Lett. {\bf 8}, 2119 (2008). 
\bibitem{Choi_2010} S.-M. Choi, S.-H.Jhi, and Y.-W. Son, Nano Lett. {\bf 10}, 3486 (2010).
\bibitem{Santos_2012} H. Santos, A. Ayuela, L. Chico, and E. Artacho, Phys. Rev. B {\bf 85}, 245430 (2012).
\bibitem{Zhang_transistor_2018} Q. Zhang, Y. Yaofeng, K. S. Chan, Z. Mu, and J. Li, App. Phys. Express {\bf 1}, 075104 (2018).
% SDW
\bibitem{Yin_NC_2016} L.-J. Yin, H. Jiang, J.-B. Qiao, and L. He, Nat. Commun. {\bf 7}, 11760 (2016).
\bibitem{Vaezi_2013} A. Vaezi, Y. Liang, D. H. Ngai, L. Yang, and E.-A. Kim, Phys. Rev. X {\bf 3}, 021018 (2013).
\bibitem{Alden_2013} J. S. Alden, A. W. Tsen, P. Y. Huang, R. Hovden, L. Brown, J. Park, D. A. Muller, and P. L McEuen, Proc. Natl. Acad. Sci. {\bf 110}, 11256 (2013).
\bibitem{San_Jose_2014} P. San-Jose, R. V. Gorbachev, A. K. Geim, K. S. Novoselov, and F. Guinea, Nano Lett. {\bf 14}, 2052 (2014).
\bibitem{Jaskolski_2020} W. Jaskolski and G. Sarbicki, Phys. Rev. B {\bf 102}, 035424 (2020).
%{Ju_Nature_2015,Pelc_2015,Lin_2013,Lane_2018}.
\bibitem{Ju_Nature_2015} L. Ju, Z. Shi, N. Nair, Y. Lv, C. Jin, J. Velasco, Jr., C. Ojeda-Aristizabal, H. A. Bechtel, M. C. Martin, A. Zettl, J. Analytis, and F. Wang, Nature (London) {\bf 520}, 650 (2015).
\bibitem{Pelc_2015} M. Pelc, W. Jaskolski, A. Ayuela, and L. Chico, Phys. Rev. B {\bf 92}, 085433 (2015).
\bibitem{Lane_2018} T. L. M. Lane, M. Andelkovic, J. R. Wallbank, L. Covaci, F. M. Peeters, and V.I. Falko, Phys. Rev. B {\bf 97}, 045301 (2018).
\bibitem{Anderson_PRB_2022} P. Anderson, Y.  Huang, Y. Fan, S. Qubbaj, S. Coh, Q. Zhou, and C. Ojeda-Aristizabal, Phys. Rev. B {\bf 105}, L081408 (2022).  
% atomic scale defects
\bibitem{Jaskolski_2016} W. Jaskolski, M. Pelc, L. Chico, and A. Ayuela, Nanoscale {\bf 8}, 6079 (2016).
\bibitem{Jaskolski_RSC_2019} W. Jaskolski and A. Ayuela, RSC Adv. {\bf 
9}, 42140 (2019).
% defected
\bibitem{Jaskolski_2019} W. Jaskolski, Phys. Rev. B {\bf 100}, 035436 (2019).
\bibitem{Jaskolski_2021} W. Jaskolski, Mol. Phys., {\bf 120}, e2013554  (2022).
\bibitem{Jaskolski_2023} W. Jaskolski, Solid State Communication {\bf 360}, 115043 (2023).
\bibitem{Pereira_2006} V. M. Pereira, F. Guinea, J. M. B. Lopes dos Santos, N. M. R. Peres, and A. H. Castro Neto, Phys. Rev. Lett. {\bf 96}, 036801 (2006).
\bibitem{Palacios_2008} J. J. Palacios, J. Fernandez-Rossier, and L. Brey, Phys. Rev. B {\bf 77}, 195428 (2008).
\bibitem{Lopez_2009} M. P. Lopez-Sancho, F. de Juan, and M. A. H. Vozmediano, Phys. Rev. B {\bf 79}, 075413 (2009).
\bibitem{Kusakabe_book} K. Kusakabe, {\it Carbon-based magnetism}, Elsevier, Amsterdam (2006).
% method
\bibitem{Gunlycke_2007} D. Gunlycke, D. A. Areshkin, J. Li, J. W. Mintmire, and C. T. White, Nano Lett. {\bf 7}, 3608 (2007).
\bibitem{Jaskolski_divac_2015} W. Jaskolski, L. Chico, and A. Ayuela, Bhys. Rev. B {\bf 91}, 165427 (2015).
\bibitem{Nardelli_1999} M. B. Nardelli, Phys. Reb. B {\bf 60}, 7828 (1999).
\bibitem{FGM} F. Garcia-Moliner and V. R. Velasco, {\it Theory of Single and Multiple Interfaces}, World Scientific, Singapore (1992).
\bibitem{Datta} S. Datta, {\it Electronic Transport in Mesoscopic Systems}, Cambridge University Press, Cambridge (1995).
\bibitem{nota_zigzag} There are two pairs of doubly-degenerate edge states (with energies $E=+V$ and $E=-V$) due to two pairs of zigzag edges of the outer layers (blue and red in Fig. \ref{fig:first}). 
Their wave functions are localized only at sublattices defined by the edge-nodes. In the investigated system the stacking is arranged in such a way that  the edge nodes of the upper half-layers are not connected to the middle layer, and therefore the corresponding edge states are not seen in the LDOS calculated in the middle layer strip. The edge-nodes of the lower half-layers are connected to the middle layer (via $t_e$), but only the left one connects directly to the sublattice defined by the vacancy.
%

\end{thebibliography}
\end{document}